\documentclass[a4paper,11pt]{article}
\pdfoutput=1 

\usepackage{jcappub} 

\usepackage{tikz}
\usepackage{amsmath}
\usepackage{amssymb}
\usepackage{xcolor}
\usepackage{subcaption}
\usepackage{float}

\usepackage[T1]{fontenc} 
\usepackage{breqn}

\usepackage{coffeestains}

\title{Quantum Recoherence in Presence of Excited States in the Early Universe}

\author[a,b, c]{Mattia Cielo,}
\author[a,b]{Simone Scarlatella,}
\author[a,b]{Gianpiero Mangano,}
\author[a,b]{Ofelia Pisanti,}
\author[a,b]{Louis Hamaide}

\affiliation[a]{INFN - Sezione di Napoli, Complesso Univ. di Monte S. Angelo, I-80126 Napoli, Italy}
\affiliation[b]{Dipartimento di Fisica ``Ettore Pancini”, Università degli studi di Napoli ``Federico II”, Complesso Univ. di Monte S. Angelo, I-80126 Napoli, Italy}
\affiliation[c]{Instituto de Física Téorica UAM/CSIC, calle Nicolás Cabrera 13-15, Cantoblanco, 28049,
Madrid, Spain}

\emailAdd{mattia.cielo@unina.it}
\emailAdd{simone.scarlatella@unina.it}
\emailAdd{gianpiero.mangano@unina.it}
\emailAdd{ofelia.pisanti@unina.it}

\abstract{We investigate the quantum-to-classical transition of primordial perturbations within a two-field inflationary framework where an adiabatic mode interacts with an entropic environment. In the case of a massive entropic environment, the attractor Bunch–Davies vacuum plays a special role: it is the only state that can undergo full recoherence, whereas all excited initial states exhibit persistent loss of purity. To characterize this behavior, we parameterize excited Gaussian initial states by their Bogoliubov coefficients and compute the purity and Rényi-2 entropy of the reduced adiabatic state as information-theoretic indicators of decoherence dynamics. We find that excited states display \emph{purity-freezing} at a non-zero plateau, where residual quantum correlations persist indefinitely, a qualitative departure from the complete recoherence observed for the Bunch–Davies vacuum. This sensitivity to initial conditions highlights the non-generic nature of full recoherence in the quantum-to-classical transition of inflationary perturbations.}

\begin{document}
\maketitle
\flushbottom


\section{Introduction}

Cosmic inflation provides a compelling framework in which quantum vacuum fluctuations are stretched and amplified by a quasi–de Sitter expansion, seeding the initial conditions of the standard cosmological model while accounting for the large-scale uniformity and the nearly scale-invariant spectrum of CMB anisotropies~\cite{Starobinsky:1980te, Sato:1981ds, PhysRevD.23.347, Linde:1981mu,Lemoine:2008zz}. 
The remarkable success of this picture supports the idea that macroscopic cosmic structures ultimately originate from microscopic quantum fluctuations. 
However, despite this intrinsically quantum origin, no direct signature of quantum correlations has yet been observed in cosmological data. 
This has motivated a broad line of research aiming to identify genuine quantum imprints in primordial perturbations and to understand how, and under what conditions, quantum signatures and coherence could survive the inflationary epoch \cite{Brandenberger:1990bx, Polarski:1995jg, Kiefer:1998qe, Burgess:2006jn, Kranas:2025jgm, Lopez:2025arw, Colas:2022hlq, Colas:2022kfu, Martin:2000xs, Martin:2021znx}.

From the theoretical viewpoint, the main challenge lies in describing the gradual emergence of classicality while preserving the fundamentally quantum dynamics of inflationary fluctuations. 
The framework of \emph{open quantum systems} (OQS)~\cite{Calzetta_Hu_2008, Lopez:2025arw} has become a natural language to address this transition \cite{Martin_2018_a, Martin_2018_b, Brahma_2022_a, Brahma_2022_b}, by explicitly tracing over unobserved degrees of freedom and describing the residual adiabatic perturbations as a reduced, generally mixed, quantum state. 
In cosmological scenarios, the system is naturally identified with the adiabatic curvature mode $\zeta$, while the environment can be associated with additional heavy or entropic fields coupled to it, or even with inaccessible sub-Hubble modes~\cite{Assassi:2013gxa}. 
The reduced dynamics captures both dissipation and decoherence, providing a consistent route to study the loss of quantum coherence in an expanding background\cite{Boyanovsky_2015, Boyanovsky_2016, Hollowood:2017bil, Burgess:2022nwu}.

Recent works have explored this framework for multifield inflationary models featuring a bilinear mixing between an adiabatic and an entropic sector, which represents the cosmological analogue of the Caldeira–Leggett paradigm~\cite{Caldeira:1981rx, Eisert_2002, Hollowood:2017bil, Colas:2022kfu, Brahma:2024ycc}. 
In particular, \cite{Colas:2022kfu} demonstrated that, in the slow-roll attractor regime, decoherence can be transient and followed by a genuine \emph{recoherence} phase, leading to a late-time recovery of purity in the adiabatic sector. 
This non-monotonic evolution has been interpreted as a consequence of the finite size and non-Markovian character of the environment. 
Subsequent analyses~\cite{Brahma:2024ycc} confirmed that such behavior is highly sensitive to the background evolution and disappears for non-attractor (ultra–slow-roll) phases, in which the growth of entanglement entropy becomes monotonic.

\medskip

In this work, we revisit this problem by focusing on the role of \emph{excited initial states} for the system, rather than on the specific form of the dynamical generator as in the time-convolutionless ($\mathrm{TCL}_2$) approach. 
While previous studies mostly assumed that the adiabatic sector starts in a Bunch–Davies configuration, we allow for generic Gaussian excitations parametrized by Bogoliubov coefficients $( \mathcal{A}_k, \mathcal{B}_k)$ that encode deviations from the standard vacuum. 
These excited configurations can arise from pre-inflationary physics, trans-Planckian effects, or departures from adiabaticity in the early stages of inflation~\cite{Brandenberger:2012aj, Ashoorioon:2014nta, Kundu:2011sg, Akama:2020jko, BouzariNezhad:2018zsi}. 
By solving the coupled system–environment dynamics exactly in a two-field setup analogous to \cite{Colas:2022kfu}, we investigate how such initial excitations affect the time evolution of purity and entropies, and whether the phenomenon of recoherence survives beyond the vacuum case.

\medskip

Our results indicate that the presence of excitations substantially modifies the late-time behavior of the reduced state. 
Even in parameter regions where full recoherence was found in the Bunch–Davies case, we observe that the purity does not return to unity. 
Instead, the system undergoes a \emph{purity freezing} at an intermediate plateau, signaling a residual level of mixedness that persists on super-Hubble scales. 
This plateau depends on both the excitation amplitude and the system–environment coupling, and provides a quantitative measure of the incomplete restoration of quantum coherence in the presence of UV-sensitive initial conditions.

\medskip

Information-theoretic quantities such as the purity and the Rényi-2 entropy provide convenient diagnostics to track this behavior. 
For Gaussian states, they can be expressed directly in terms of the covariance matrix of the adiabatic mode and allow a transparent comparison between decoherence, recoherence, and freezing regimes. 
We will show that, although the dynamics remains Gaussian and fully unitary in the whole system, the reduced adiabatic sector never achieves complete recoherence once excitations are present.

\medskip

The paper is organized as follows. 
In Sec.~\ref{sec:model} we review the two-field model and set up the dynamical equations for the coupled adiabatic–entropic system. 
Sec.~\ref{sec:purity} introduces the information-theoretic framework used to quantify mixedness and entanglement through purity, Rényi-2 and Von Neumann entropies. 
In Sec.~\ref{sec:results} we present the numerical results for the evolution of these quantities in the presence of excited initial states, highlighting the emergence of the purity-freezing plateau and its dependence on the excitation parameters. 
We conclude in Sec.~\ref{sec:conclusions} with a discussion of the implications for quantum signatures in cosmological perturbations.

\section{Two-field effective model in (quasi-)de Sitter} 
\label{sec:model}
Data retrieved in inflationary cosmology experiments, in particular observations of the cosmic microwave background (CMB), provide evidence that the primordial spectrum is mostly sourced by adiabatic fluctuations while isocurvature modes are highly suppressed \cite{Planck:2018vyg}. It was then shown that a possible way to characterize the dynamics of the curvature perturbations was to consider a two-field model, in order to account for both the adiabatic and the isocurvature modes evolution in the early universe. 
This model has turned out to be extremely useful from a phenomenological point of view as well as a perfect framework for studying the long-standing problem of the classicalization of primordial perturbations. 

In this scheme, the \textit{system} couples at leading order to a hidden entropic field, the \textit{environment} on a (quasi-)de~Sitter background with scale factor $a(\eta)$ in conformal time $\eta$ \cite{Starobinsky:1980te,Starobinsky:1992ts}. 

The model is Gaussian and features minimal bilinear mixing that preserves spatial homogeneity and isotropy while enabling energy/information exchange between the the system (adiabatic) and environment (entropic) sectors. 

\subsection{Lagrangian in the $\zeta -\mathcal{F}$ variables}

The effective Lagrangian density considered is \cite{Assassi:2013gxa}
\begin{equation}
\mathcal{L} = 
a^{2}\epsilon M_{\rm Pl}^{2}(\zeta')^{2} 
- a^{2}\epsilon M_{\rm Pl}^{2}(\partial_{i}\zeta)^{2} 
+ \tfrac{1}{2}a^{2}(\mathcal{F}')^{2} 
- \tfrac{1}{2}a^{2}(\partial_{i}\mathcal{F})^{2} 
- \tfrac{1}{2}\mu^{2}a^{4}\mathcal{F}^{2} 
+ \lambda a^{3}\sqrt{2\epsilon}\,M_{\rm Pl}\,\zeta' \mathcal{F} .
\label{eq:L-zeta-F}
\end{equation}
where $\epsilon \equiv -\dot H/H^{2}$ is the first Slow-Roll (SR) parameter \footnote{As usual, dot and prime denote derivatives with respect to cosmic time $t$ and conformal time $\eta$, respectively.}, $M_{\rm Pl}$ is the reduced Planck mass, $\mu$ the mass of the environment field, and $\lambda$ is a dimensionful interaction parameter.

The $\zeta$ field represents the adiabatic perturbation, from now on the ``system'', while $\mathcal{F}$ field is the environment with which the system interacts, leading to a decoherence process.

It is convenient to introduce the canonical Mukhanov--Sasaki fields by scaling both fields, in order to obtain for each of them a Sasaki-like equation, 
\begin{equation}
z^{2} \;\equiv\; 2 a^{2} \epsilon M_{\rm Pl}^{2}\,, 
\qquad v \;\equiv\; z\,\zeta\,,
\qquad u \;\equiv\; a\,\mathcal{F}\,,
\label{eq:canon-defs}
\end{equation}
for which the free parts of Eq.~\eqref{eq:L-zeta-F} take their standard canonical form, and the background “squeezing” terms are encoded in $z''/z$ and $a''/a$ . Using $\zeta'=(v'/z) - (z'/z^{2})v$ and $F=u/a$, the mixing term becomes
\begin{equation}
\mathcal{L}_{\rm mix} \;=\; \lambda\,a\;\Big(v' - \frac{z'}{z}\,v\Big)\,u\,,
\label{eq:Lmix-vu}
\end{equation}
so that the total quadratic action reads
\begin{eqnarray}
S &=& \tfrac{1}{2} \!\int\! d\eta\, d^{3}x \left[ 
\left(v' - \tfrac{z'}{z}v\right)^{2} - (\nabla v)^{2} 
+ \left(u' - \tfrac{z'}{z}u\right)^{2} - (\nabla u)^{2} 
- \mu^{2}a^{2}u^{2} \right. \nonumber \\
&+& \left. \lambda a \left(v' - \tfrac{z'}{z}v\right)u 
\right].
\label{eq:Sv-u}
\end{eqnarray}

The mixing is linear in $v'$ and $v$, and thus,  the theory remains Gaussian and amenable to exact mode-by-mode analyses and to open-system reductions (e.g., via TCL master equations) \cite{Colas:2022hlq,Colas:2022kfu,Brahma:2024yor,Brahma:2024ycc}. 

The conjugate momenta are
\begin{equation}
\pi_S \;\equiv\; \frac{\partial \mathcal{L}}{\partial v'} \;=\; v' -\frac{z'}{z}v+ \lambda\,a\,u\,,
\qquad
\pi_E \;\equiv\; \frac{\partial \mathcal{L}}{\partial u'} \;=\; u' - \frac{a'}{a}u\,,
\label{eq:conj-momenta}
\end{equation}
where the shift in $\pi_S$ is due to the derivative mixing in Eq.~\eqref{eq:Lmix-vu}. The Hamiltonian density, $\mathcal{H} \equiv \pi_S v' + \pi_E u' - \mathcal{L}$, can be written as
\begin{equation}
\mathcal{H} = \tfrac{1}{2}\!\Big[\pi_S^{2} - (\partial_i v)^{2} + \tfrac{z'}{z}\{\pi_S,v\}\Big] 
+ \tfrac{1}{2}\!\Big[\pi_E^{2} + (\partial_i u)^{2} + (\mu^{2}+\lambda^{2})a^{2}u^{2} + \tfrac{a'}{a}\{\pi_E,u\}\Big] 
- \lambda a\,\pi_S u,
\label{eq:H-density}
\end{equation}
which is quadratic and mode-separable in Fourier space. This form makes transparent both the (shifted) kinetic structure induced by the mixing and the background-induced effective mass. In order to quantize the theory we use the interaction picture, so that the equations of motion for the two rescaled fields are two uncoupled Mukahnov-Sasaki differential equations,
\begin{align}
v_{k}'' + \Big(k^{2} - \frac{z''}{z}\Big)\,v_{k}
&\;=\; 0\,,
\label{eq:eom-vk}\\[4pt]
u_{k}'' + \Big(k^{2} + m^{2} a^{2} - \frac{a''}{a} \Big)\,u_{k}
&\;=\; 0\,,
\label{eq:eom-uk}
\end{align}
where $m^2=\mu^2+\lambda^2$ is the effective mass, $a = -1/(H \tau)$ and $z^2(\tau) = 2 a^2 \epsilon M^2_{Pl} $ , \cite{MUKHANOV1992203, Mukhanov:2007zz}. 

These two equations admit the following two expressions as mode functions:
\begin{align}
f_{k}^{(S)}(\eta) &\;=\; \frac12 e^{i \frac{\pi}{2} (\nu_S + \frac12)} \sqrt{-\pi \eta} \, H^{(1)}_{\nu_S}(-k\eta \,) \,, \qquad \nu_S^2=\frac94 \,,
\label{eq:eom-solvk}\\[4pt]
f_{k}^{(E)}(\eta) &\;=\; \frac12 e^{i \frac{\pi}{2} (\nu_E + \frac12)} \sqrt{-\pi \eta} \, H^{(1)}_{\nu_E}(-k\eta \,) \,, \qquad \nu_E^2=\frac94 - \frac{m^2}{H^2} \,.
\label{eq:eom-soluk}
\end{align}

Given the solutions (\ref{eq:eom-solvk}) and (\ref{eq:eom-soluk}), we can construct the most general solution of Eq.s (\ref{eq:eom-vk}) and (\ref{eq:eom-uk}) as
\begin{equation}
v_k = \mathcal{A}_k^{(S)} f_k^{(S)} + \mathcal{B}_k^{(S)} f_k^{(S)*}, ~~~~~~ u_k = \mathcal{A}_k^{(E)} f_k^{(E)} + \mathcal{B}_k^{(E)} f_k^{(E)*},
    \label{general_sol}
\end{equation}
where \(\mathcal{A}_k\) and \(\mathcal{B}_k\) are the Bogoliubov coefficients which define the initial condition of the system and environment. We allow for excited initial states for the system, i.e. $|\mathcal{B}_k^{(S)}| \neq 0 $, which quantifies the deviation from the Bunch-Davies vacuum. This choice is physically motivated by several considerations. 

Indeed, inflation is known to be past-incomplete, requiring a pre-inflationary phase whose dynamics can leave observable imprints on the initial quantum state. Unlike deep subhorizon environmental modes, which remain far inside the horizon throughout the relevant dynamics, the system modes of cosmological interest cross the horizon during inflation. These modes are therefore sensitive to trans-Planckian physics or any preceding cosmological phase, which can naturally populate excited states prior to horizon crossing, \cite{Cielo:2022vmo, Cielo:2023enz, Cielo:2024poz, Borde:2001nh, Martin:2000xs, BouzariNezhad:2018zsi, Kempf:2000ac, Holman:2007na, Brahma:2019unn, Ashoorioon:2010xg, Ashoorioon:2014nta}. 

Furthermore, modes near horizon crossing have characteristic energies \(E_{\text{sys}} \sim aH\), which are not exponentially suppressed when compared with the de Sitter temperature \(T = H/(2\pi)\). In contrast, environmental modes deep inside the horizon have energies \(E_{\text{env}} \sim k \gg aH\), leading to Boltzmann suppression factors that render excitations negligible. This hierarchy in energy scales justifies treating the environment as remaining in its vacuum state, while allowing for non-trivial initial conditions for the observable perturbations, \cite{Lopez:2025arw}.

Recent phenomenological analyses also suggest that modest initial state excitations can improve consistency with observational data, providing additional empirical motivation for considering such configurations, see \cite{Brahma:2025dio, Ashoorioon:2014nta, Planck:2018vyg}.

It is worth recalling that the Bogoliubov coefficients satisfy general consistency conditions following from the conservation of the Wronskian associated with the mode equation (or, equivalently, from the canonical commutation relations of the quantum field operators). This requirement leads to
\begin{equation}
|\mathcal{A}_k|^2 - |\mathcal{B}_k|^2 = 1 \, .
\end{equation}
This relation is particularly useful for introducing a convenient parametrization of the two Bogoliubov coefficients. A simple and widely used choice is inspired by the so-called \emph{alpha-vacua}, where one defines
\begin{equation}
\mathcal{A}_k = \cosh{\alpha} \,, \qquad 
\mathcal{B}_k = e^{i\theta}\sinh{\alpha} \, ,
\end{equation}
with $\alpha$,$\theta$ real free parameters. We set the phase $\theta=0$ and work only with $\alpha$, which may take both positive and negative values \cite{Cielo:2024poz, Gong:2023kpe, Aravind:2013lra, Kanno:2022mkx, Allen:1985ux}\footnote{See also~\cite{Danielsson:2002kx, Broy:2016zik}.}.

Moreover, the $\alpha$-vacua can be interpreted as \emph{squeezed states} with respect to the Bunch--Davies vacuum, implying that such excited states remain Gaussian \cite{Mukhanov:2007zz}\footnote{This property is crucial, as it ensures the consistency of the master equation and of the open effective field theory formalism for quantum perturbations during inflation, both of which rely on the Gaussian character of the system’s density matrix. On top of that, one can easily check that the Green's matrix for the system, in the Open Effective approach is left completely unchanged upon the replacement of the general solution by mean of the Bogoliubov coefficients, since, after the computation, $ \tilde{G}_{\alpha} = \textbf{Re}(|\mathcal{A}_k|^2 - |\mathcal{B}_k|^2 ) G$, see the Supplementary Material in \cite{Colas:2022kfu}.}.

In the remainder of this article, we will analyse how the presence of excitations in the primordial mode functions influences the evolution of the quantum indicators introduced in the following sections, namely, the purity and two types of entropy measures that are particularly useful in the study of open quantum systems: the entanglement entropy and the Rényi-2 entropy. 

Before concluding this section, it is important to note that although the parameter $\alpha$ can, in principle, take any real value, its magnitude is directly related to the squeezing parameter that appears in the final expression for the primordial power spectrum, making these quantities observable~\cite{Brahma:2025dio, Broy:2016zik, Ashoorioon:2014nta}. To preliminarily constrain the parameter space, we must ensure that the average energy density associated with adiabatic perturbations does not exceed the total background energy density, namely, that the inflaton potential remains the dominant contribution during inflation, as discussed in~\cite{Cielo:2024poz}. 

For this reason, we will restrict the analysis to the range $-1 \leq \alpha \leq 1$. 

To summarise, our model will thus involve three free parameters: the effective mass parameter,
$m$, which characterises the mass of the isocurvature field evolving alongside the adiabatic perturbations (and acts as a reservoir dynamically coupled to the adiabatic sector); the coupling strength between the two sectors, $\lambda$; and finally, the parameter $\alpha$, which quantifies the impact of the initial conditions in the adiabatic sector. 

As it is clear from the index in Eq. (\ref{eq:eom-soluk}), in the case of a very massive environmental field, i.e. $\frac{m^2}{H^2} \geq 9/4$, the index $\nu_{E}$ can take imaginary values. This will be precisely one of the cases with the most interest in light of the fact that this regime can exhibit a full recoherence behaviour for the inflationary quantum perturbations, as described in \cite{Colas:2022kfu}.

\subsection{Evolution of the environment-system set: The Transport Method}

In this section, we will briefly review the numerical strategy to solve the full dynamics for both the system and the environment sectors. 
In this work, we will set our description for the evolution of the perturbation in a perfect deSitter spacetime within the Slow-Roll approximation. For extension to Ultra Slow Roll cases (USR), see for example \cite{Brahma:2024ycc}.  
The Lagrangian (\ref{eq:L-zeta-F}) enables us to solve the dynamics for the full theory, encompassing both the system and its environment. We will make use of the Transport Equations Method (TEM), firstly developed in \cite{Seery:2012vj} and then applied in the context of an open quantum system in the early universe in \cite{Brahma:2024ycc,Colas:2022kfu,Colas:2022hlq}. 

In order to show our numerical setup, we note that the theory described in the previous section can be recast into a matrix form. So, defining the whole Hamiltonian for the system to be
\begin{equation}
H(\eta) =
\begin{pmatrix}
H_{(S)}(\eta) && H_{\mathrm{int}}(\eta) \\[4pt]
H_{\mathrm{int}}^{T}(\eta) && H_{(E)}(\eta)
\end{pmatrix},
\label{eq:S5}
\end{equation}
we can divide the three sectors of the theory into two free ones ($H_{(S)}$ and $H_{(E)}$) and the interaction one $H_{int}$:
\begin{align}
H_{(S)}(\eta) &=
\begin{pmatrix}
\dfrac{k^{2}}{a^{2}} && \dfrac{a'}{a} \\[4pt]
\dfrac{a'}{a} && 1
\end{pmatrix},
&
H_{(E)}(\eta) &=
\begin{pmatrix}
k^{2} + (m^{2} + \lambda^{2})a^2 && \dfrac{a'}{a} \\[4pt]
\dfrac{a'}{a} && 1
\end{pmatrix},
&
H_{\mathrm{int}}(\eta) &=
\begin{pmatrix}
0 && 0 \\[4pt]
-\lambda a && 0
\end{pmatrix} \, ,
\label{eq:S6}
\end{align}
and introduce a phase-space vector with the field variables, $q^T = (v , p_v , u, p_u)$. Within this formalism, the dynamics of the system is obtained from the covariance matrix, by tracing over the environment. With the reduced density matrix $\Tilde{\rho}(\eta) = \text{Tr}_{\mathrm{E}}(\rho_{\mathrm{S}}(\eta) \otimes \rho_{\mathrm{E}}(\eta_0))$\footnote{Since there is no natural choice of the initial density matrices for the system or environmental fields, it is typically assumed that the state of the environment is stationary due to the choice of typical timescales $\eta_{\mathrm{E}} \gg \eta_{\mathrm{S}}$ (Born approximation), which makes the environment effectively act as a reservoir that exchanges energy and information with the system.} being Gaussian at initial conditions, the reduced covariance for the system reads:
\begin{equation}
    \Sigma^{(S)}_{ij}(\eta) \equiv \frac12\text{Tr}_{\rm S} \Big[ \{q_i(\eta),q_j^\dagger(\eta)\} \Tilde{\rho}(\eta_0) \Big] \, \quad\quad \rm with \quad i,j=1,2 .
    \label{eq:covmatrix}
\end{equation}

The observables are associated with the elements $\Sigma_{ij}$, and these fully characterize a Gaussian state (even in the presence of initial excited states).

However, it can be show that the following complete set of transport equations can be derived for the total unitary evolution:
\begin{equation}
    \frac{d}{d\eta}\Sigma^{(\rm S +E)} = 2 \Omega H^{(\rm S +E)}\Sigma^{(\rm S +E)} - 2\Sigma^{(\rm S +E)}H^{(\rm S +E)} \Omega \, ,
    \label{eq:exacttm}
\end{equation}
where $\Omega$ is a block-diagonal matrix, $\Omega = \bigl(\begin{smallmatrix} 0 & 1 \\ -1 & 0 \end{smallmatrix}\bigr)$ , with antisymmetric entries and $H^{(\rm S+E)}$ is the Hamiltonian of the full system depicted above.

In our numerical simulations, we add to Eqs. (\ref{eq:exacttm}) an extra equation for the determinant of the system, which will read as: 
\begin{equation}
    \frac{d}{d\eta}\text{det}\,\Sigma^{(\rm S +E)} = \Sigma_{11}^{(\rm S +E)} \frac{d \Sigma^{(\rm S +E)}_{22}}{d\eta} +\Sigma_{22}^{(\rm S +E)} \frac{d \Sigma^{(\rm S +E)}_{11}}{d\eta} -2\Sigma_{12}^{(\rm S +E)} \frac{d \Sigma^{(\rm S +E)}_{12}}{d\eta} \,.
\end{equation}
In principle, this equation is redundant, and one can derive $\text{det}\,\Sigma$ after having determined the solutions of Eq.(\ref{eq:exacttm}). However, in order to compute the purity, since there are cancellations between quantities that diverge at late time, it is numerically more efficient to treat it as an independent quantity in the differential equation system.
In this way we are left with eleven differential equations that we solve numerically, using the number of e-folds, $N$, as the independent variable. In particular, $N$ is defined as the number of e-folds for a mode with $k=1$, that crosses the horizon at $N=0$. We set our initial conditions well inside the horizon, $N_{in}= -10$, until $N_{fin} = 10$, after it crosses the horizon. In our treatment, we also consider only dimensionless model parameters relative to a constant Hubble function (we set its value to be $H=1$ in our numerical simulation)\footnote{Taking a constant value for H is one of the main approximations used in the literature to compute the quantum markes for a single mode $k$. However, in order to make a precise evaluation of the resummed effects to the power spectrum, for example, one would need to compute such evolution for a set of modes $k$ and accounting for an evolution of the $H$ functions over time. We leave this analysis for a future work.}. 

As we said in the previous section, our goal is to see the effect of evolving the system sector from a non Bunch-Davies state. For that, we need to compute the initial conditions for the initial correlators of the fields in the phase space. Using the Bogoliubov mode functions, we can rewrite the initial conditions at the lowest order in the $\eta$ expansion around $- \infty$ for the system field and conjugate momentum as: 
\begin{eqnarray}
\Sigma_{11}^{(\text{in})} &=& \frac{\cosh (2 \alpha )+\sinh (2 \alpha ) \cos (2 k \eta )}{2 k}, \\
\Sigma_{12}^{(\text{in})} &=& -\frac{1}{2} \sinh (2 \alpha ) \sin (2 k \eta ), \\
\Sigma_{22}^{(\text{in})} &=& \frac{k}{2} \left[ \cosh (2 \alpha) - \sinh (2 \alpha ) \cos (2 k \eta )\right].
\end{eqnarray}
One can easily see that, at this order and for $\alpha = 0$, we recover exactly the initial conditions for the free field in the Bunch-Davies case.

The corresponding initial conditions for the covariance elements of the environment, valid for the BD case and matching the free Hankel solutions at early times, are
\begin{equation}
\Sigma_{33}^{(\text{in})} = \frac{1}{2k},\qquad \Sigma_{34}^{(\text{in})} = 0,\qquad \Sigma_{44}^{(\text{in})} = \frac{k}{2}.
\end{equation}

\section{Quantum information measures}
\label{sec:purity}
A useful way to characterize the quantum-to-classical transition of cosmological perturbations
is through the purity and the entropic content of the reduced state, describing the adiabatic mode
after tracing out the environment \cite{Eisert_2002, Boyanovsky_2018, Brahma_2022_a}.
As already mentioned, for a Gaussian state, the system reduced density matrix $\rho_S$
is completely determined by its covariance matrix $\Sigma^{(S)}_{ij}$. 
The \emph{purity} is defined as
\begin{equation}
    \gamma \equiv \mathrm{Tr}\big(\rho_S^2\big)
    = \frac{1}{4\,\det \Sigma^{(S)}}.
    \label{eq:purity}
\end{equation}
A pure state corresponds to $\gamma=1$, whereas $\gamma<1$ signals the emergence of decoherence.
As shown in Refs. \cite{Colas:2022kfu} , the dynamics of $\gamma$ in de~Sitter backgrounds
is non-monotonic: after an initial decay (decoherence), a turning point can appear, leading to a partial
\emph{recoherence} when the effective coupling to the environment weakens on super-Hubble scales.

\medskip

Another interesting quantum informatic measure is the von Neumann entanglement entropy of the system density matrix,
\begin{equation}
    S_{\mathrm{ent}} = -\mathrm{Tr}\,(\rho_S\ln\rho_S),
\end{equation}
which, for a single Gaussian mode, can be written in terms of the
occupation number $n_k$ as
\begin{equation}
    n_k = \frac{1-\sqrt{\gamma}}{2\sqrt{\gamma}},
    \qquad
    S_{\mathrm{ent}} = 2\Big[(1+n_k)\ln(1+n_k) - n_k\ln n_k\Big].
    \label{eq:ent_entropy}
\end{equation}
In this form, $S_{\mathrm{ent}}$ grows as the system loses coherence,
while its monotonic decrease (equivalently, increasing $\gamma$)
signals partial recoherence.
Finally, an equivalent and often more convenient quantifier of mixedness is provided by the
Rényi-2 entropy,
\begin{equation}
    S_2(\rho_S) = -\ln \mathrm{Tr}\,(\rho_S^2) = -\ln \gamma,
    \label{eq:renyi2}
\end{equation}
which enjoys several remarkable properties for Gaussian states and provides a direct and computationally robust proxy for the von~Neumann entropy in dynamical Gaussian settings.
It satisfies the strong subadditivity condition,
is expressible in closed form through the covariance determinant,
and provides a consistent operational measure of correlations, decoherence,
and information exchange between system and the environment.
For these reasons, $S_2$ plays the role of a natural entropy in open quantum
field systems and constitutes the backbone of Gaussian quantum information theory
\cite{PhysRevLett.109.190502, Adesso_2012}.
Within a thermodynamic interpretation, the family of Rényi-$q$ entropies
\begin{equation}
    S_q(\rho_S) = \frac{1}{1-q}\ln\mathrm{Tr}\,(\rho_S^q)
\end{equation}
interpolates between the purity ($q=2$) and the von~Neumann limit ($q\to1$),
and can be related to variations of the free energy between two effective
temperatures $T$ and $T'$ with $q=T/T'$~\cite{Baez:2011upp}.
In this sense, $S_q$ represents a lower bound on entropy production,
with $S_2$ being a conservative measure of decoherence.

In recent years, generalized entropy measures have attracted growing interest. On the one hand, beyond the thermodynamic aspects mentioned earlier  \cite{Kirchanov:2008,Baez:2011upp,Lavagno_2002}, Rényi entropy has acquired an interpretative role in theoretical contexts, revealing universal scaling laws in QFT, with specific applications in CFT and in black hole physics, by comparing the Bekenstein entropy with respect to the Rènyi's one \cite{kusuki2025universalityrenyientropyconformal,Kudler_Flam_2023,Tsallis_2013,abreu2025modifiedtsallisrenyientropymondlike,Czinner_2025}. On the other hand, Rényi-$q$ entropies have proven to be valuable tools in measure theory, increasingly establishing themselves as benchmarking instruments for quantum algorithms and devices, with numerous applications in many-body physics models and lattice implementations \cite{PhysRevLett.109.190502,Brydges_2019,ozawa2024perspectivephysicalinterpretationsrenyi,Elben_2019}. Thus, interest in generalized entropies lies both in their diagnostic utility and in their physical interpretations.

In the following, we shall use the purity $\gamma$ and its Rényi-2 counterpart
$S_2(\rho_k)=-\ln\gamma$ to trace the evolution of quantum coherence in the
presence of excited initial states along the standard Von-Neumann definition.

\section{Numerical Solutions}
\label{sec:results}

\begin{figure}[t]
    \centering
    \begin{subfigure}[b]{0.85\linewidth}
        \centering
        \includegraphics[width=\linewidth]{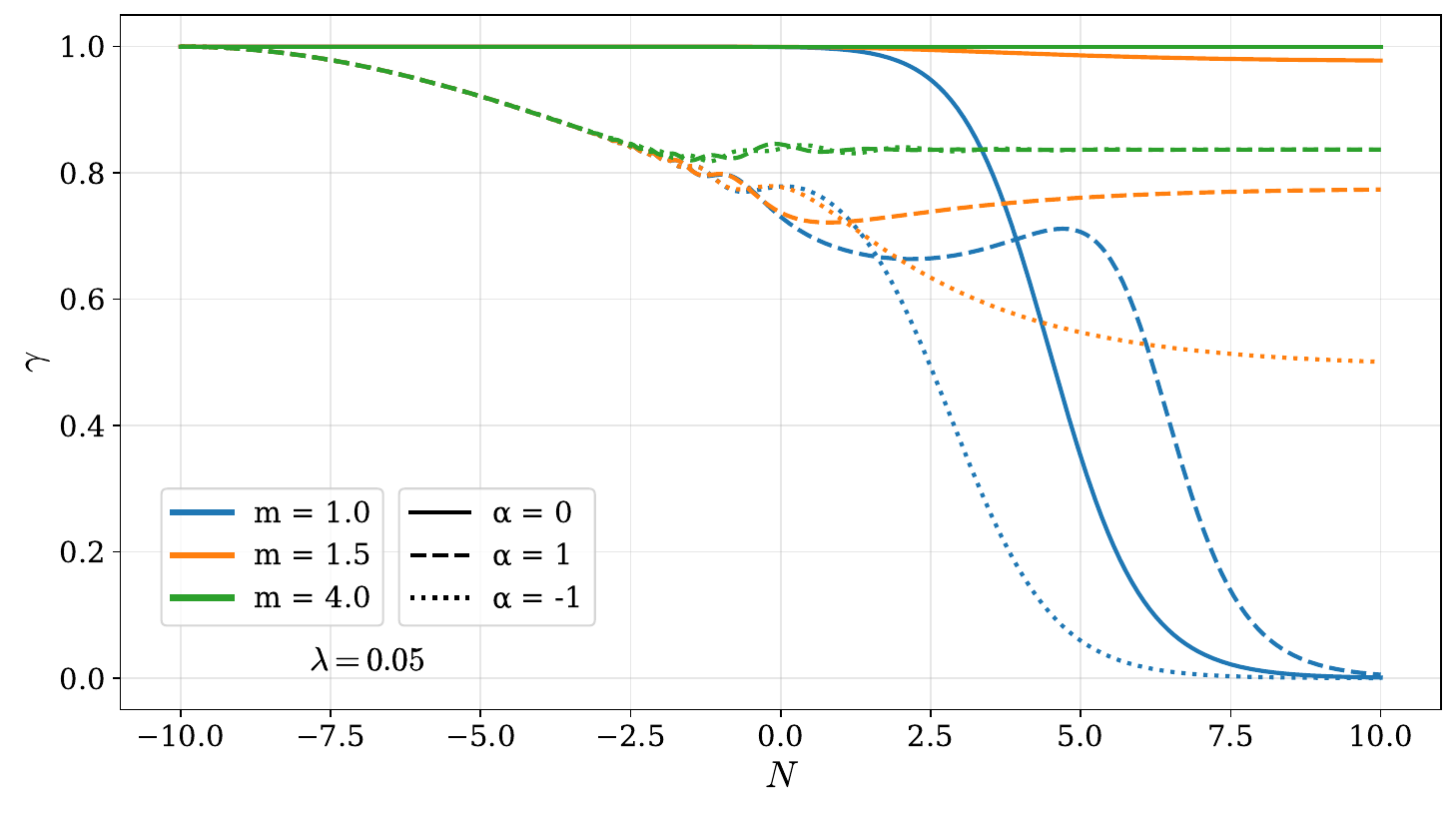}
        \caption{Weak coupling $\lambda=0.05$. For $m/H=\{1,1.5\}$ the BD case ($\alpha=0$) shows a monotonic decoherence, while for $m/H=4$ reproduce the non-monotonic vacuum trend (decoherence followed by recoherence) consistent with the benchmark; for $\alpha=\pm 1$ a small super-horizon bump precedes a delayed but complete decoherence for $m/H =1$ while purity freezing in all other cases.}
        \label{fig:purity-weak}
    \end{subfigure}
    \vspace{0.4cm}
    \begin{subfigure}[b]{0.85\linewidth}
        \centering
        \includegraphics[width=\linewidth]{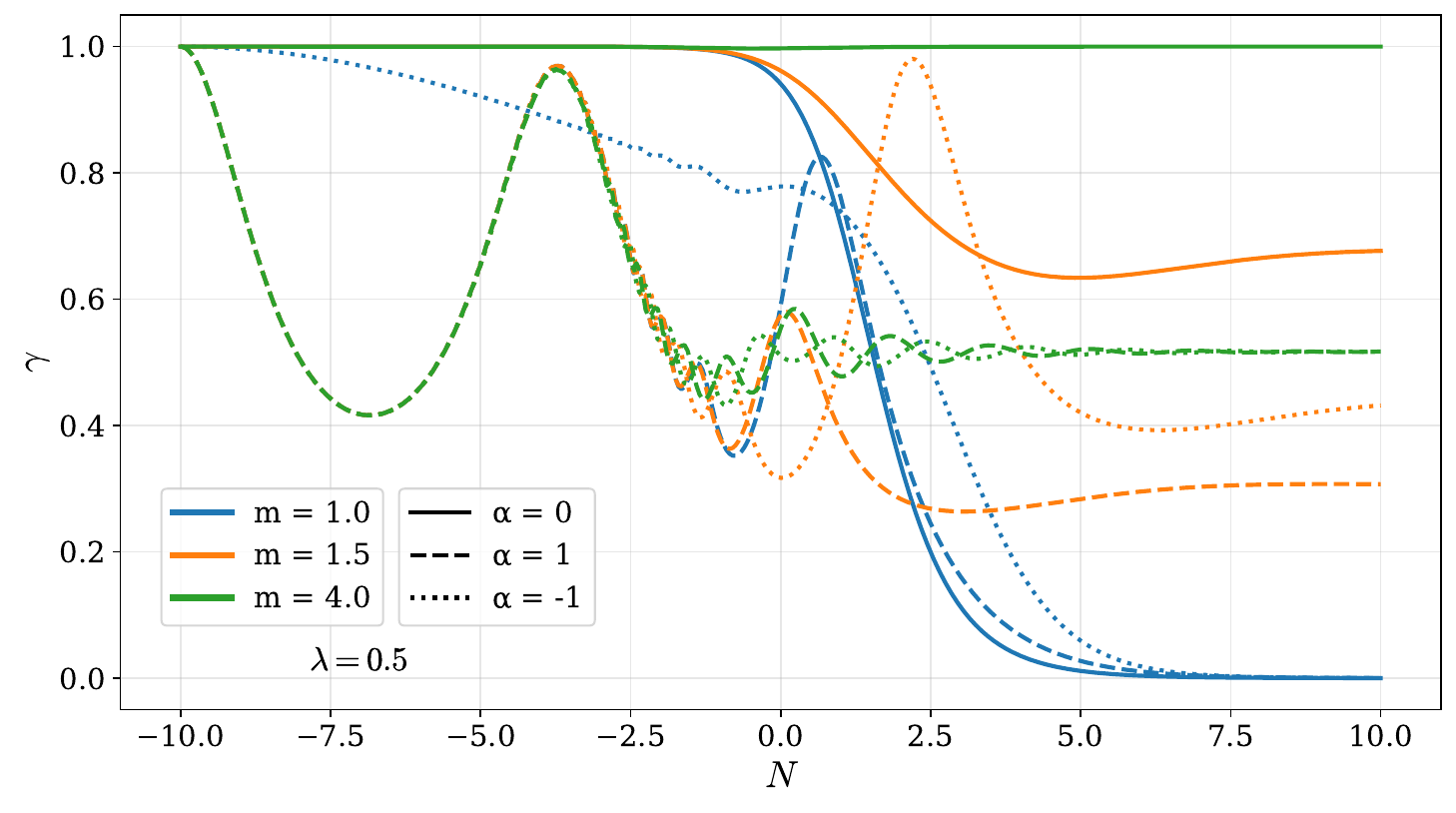}
        \caption{Strong coupling $\lambda=0.5$. Sub-horizon oscillations in $\gamma$ appear before $N_{\rm cr}=0$ and are enhanced for lighter environments; the BD case retains the vacuum recoherence signature for $m/H=4$, while $\alpha=\pm 1$ display a transient super-horizon rebound followed by a purity freezing at late times.}
        \label{fig:purity-strong}
    \end{subfigure}
    \caption{Purity evolution $\gamma(N)$ for $\alpha\in\{-1,0,1\}$ and $m/H\in\{1,1.5,4\}$.}
    \label{fig:purity-scan}
\end{figure}

\begin{figure}[t]
    \centering
    \includegraphics[width=0.82\linewidth]{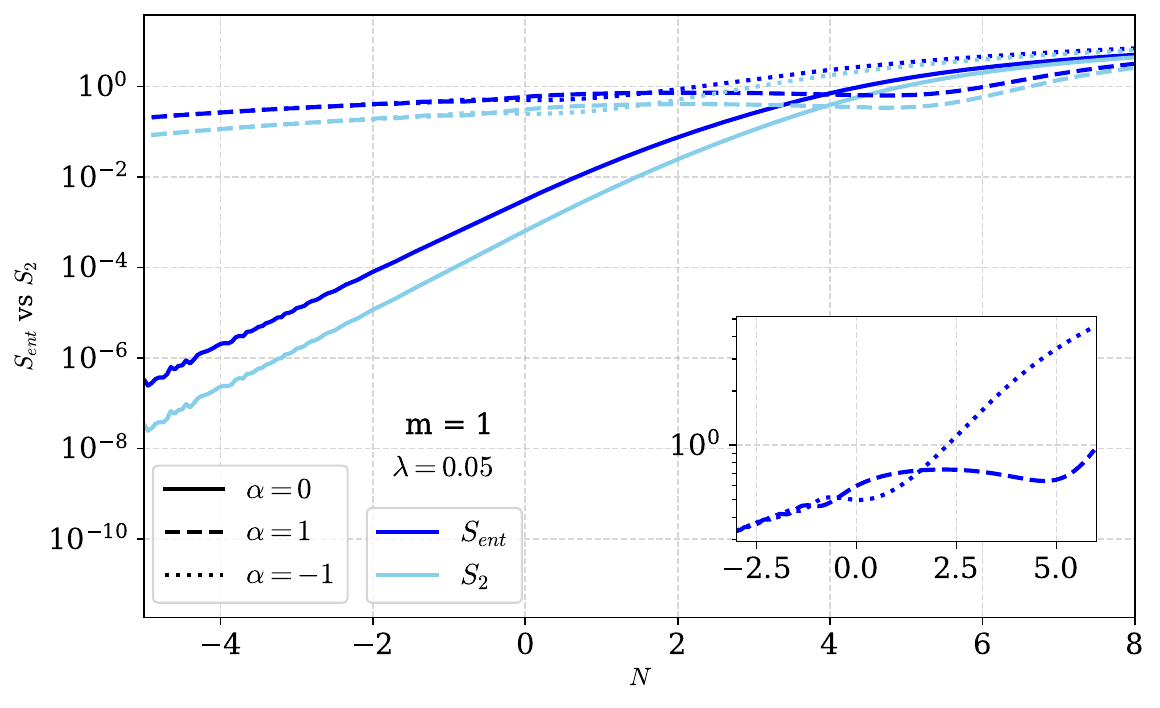}
    \caption{Weak coupling $\lambda=0.05$, $m/H=1$. Rényi-2 $S_2$ stays below von Neumann $S_{\rm ent}$ and tracks its time dependence; for $\alpha=0$ both entropies grow monotonically after crossing, while for $\alpha=\pm 1$ a small super-horizon dip accompanies the transient purity rebound, followed by sustained entropy growth at late times.}
    \label{fig:entropy-weak-m1}
\end{figure}

\begin{figure}[t]
    \centering
    \includegraphics[width=0.82\linewidth]{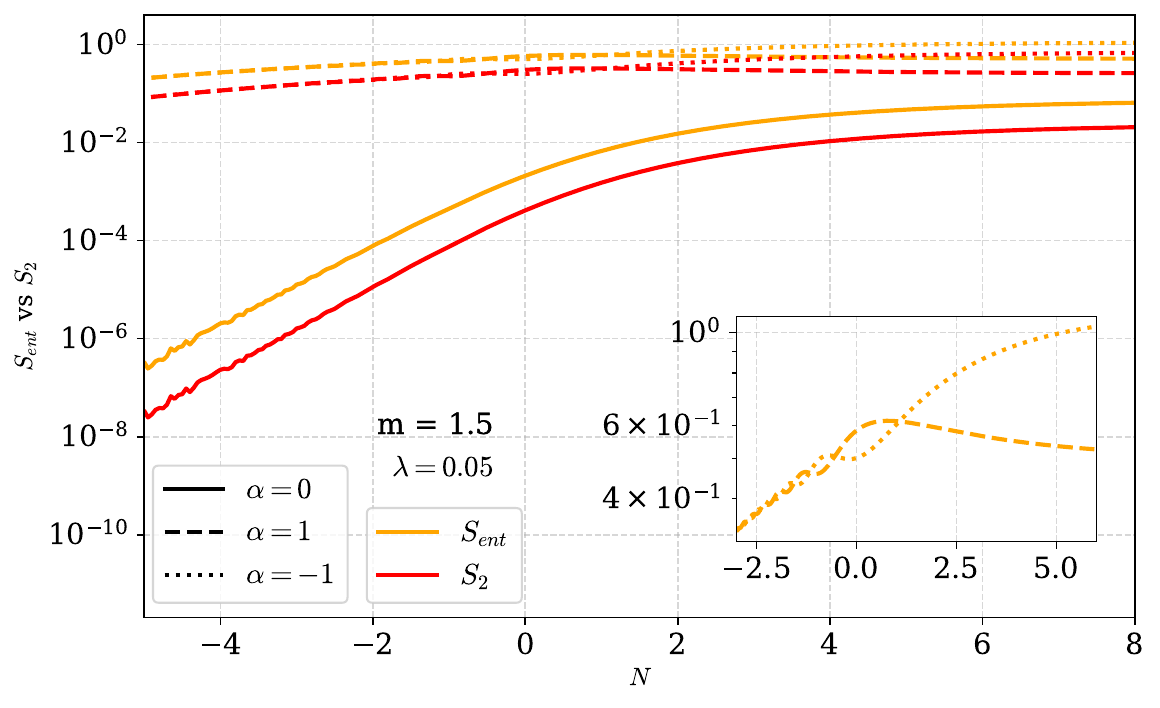}
    \caption{Weak coupling $\lambda=0.05$, $m/H=1.5$. The overall behavior mirrors the $m/H=1$ case with slightly reduced growth rates; excited initial states $\alpha=\pm 1$ again show a shallow early dip before converging to larger late-time entropy than the BD trajectory.}
    \label{fig:entropy-weak-m15}
\end{figure}

\begin{figure}[t]
    \centering
    \includegraphics[width=0.82\linewidth]{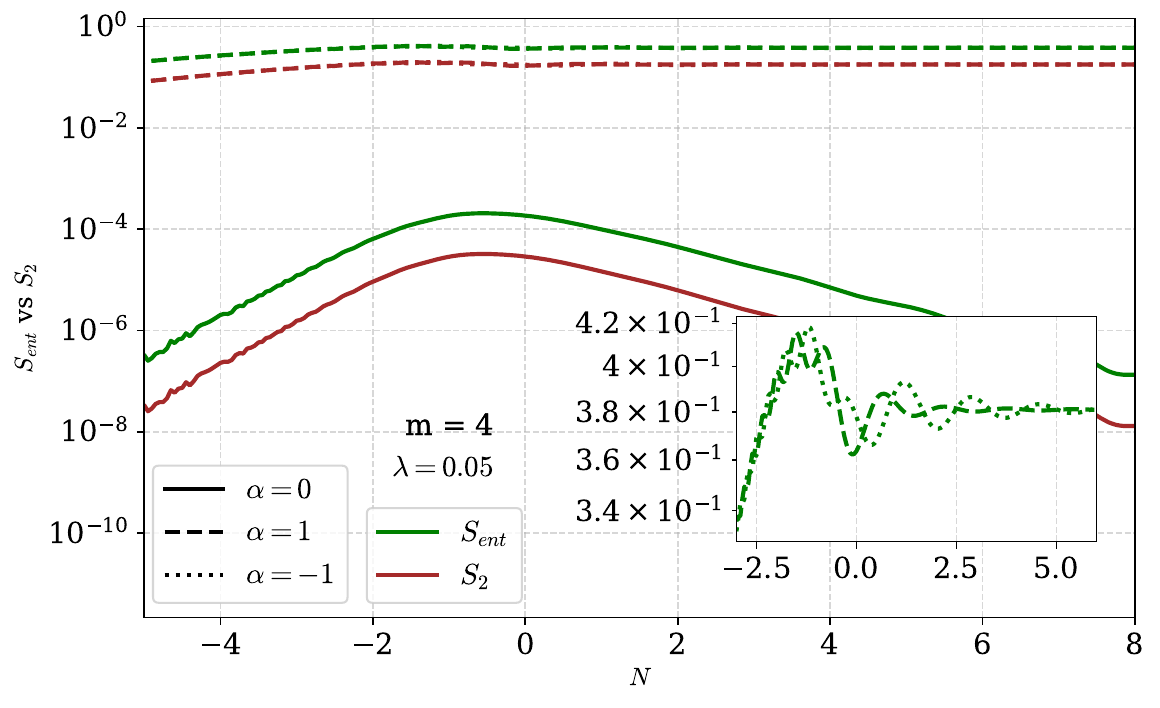}
    \caption{Weak coupling $\lambda=0.05$, $m/H=4$. The BD vacuum ($\alpha=0$) displays the non-monotonic trend with an eventual decrease toward zero entropy, whereas $\alpha=\pm 1$ exhibits only a transient reduction, consistent with a purity freezing scenario.}
    \label{fig:entropy-weak-m4}
\end{figure}

\begin{figure}[t]
    \centering
    \includegraphics[width=0.82\linewidth]{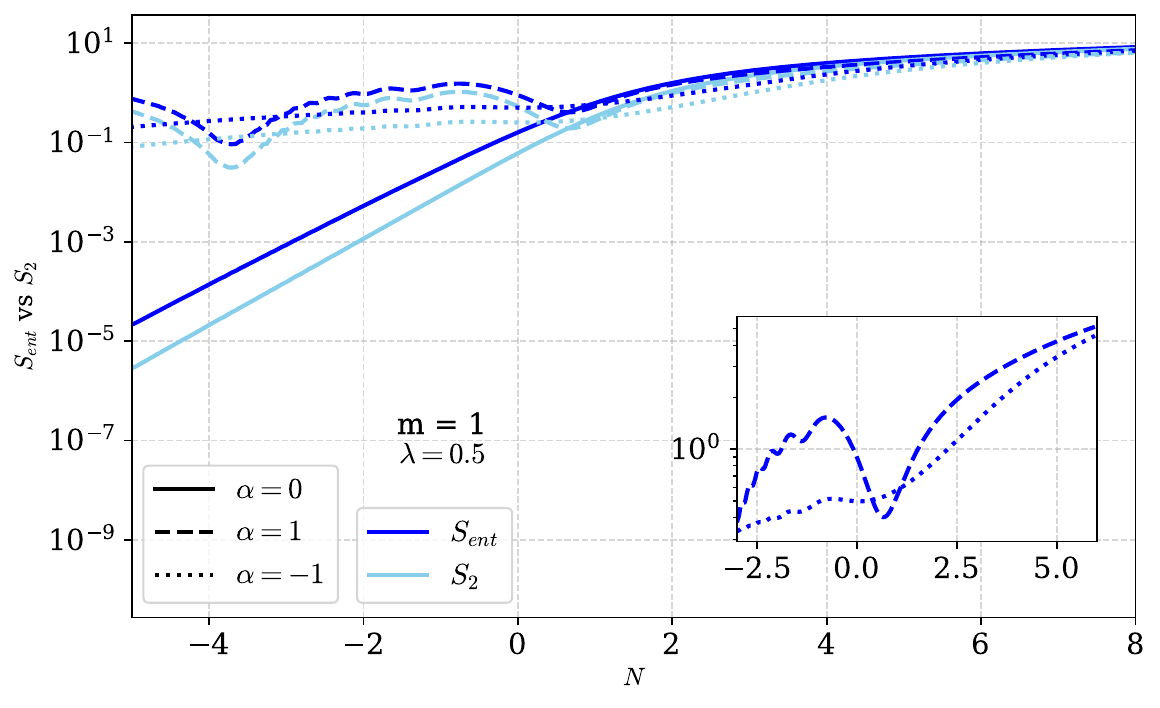}
    \caption{Strong coupling $\lambda=0.5$, $m/H=1$. Sub-horizon oscillations imprint small ripples before $N_{\rm cr}=0$; post-crossing, both $S_{\rm ent}$ and $S_2$ grow faster than in the weak-coupling case, with excited states showing a transient dip then sustained increase toward late times.}
    \label{fig:entropy-strong-m1}
\end{figure}

\begin{figure}[t]
    \centering
    \includegraphics[width=0.82\linewidth]{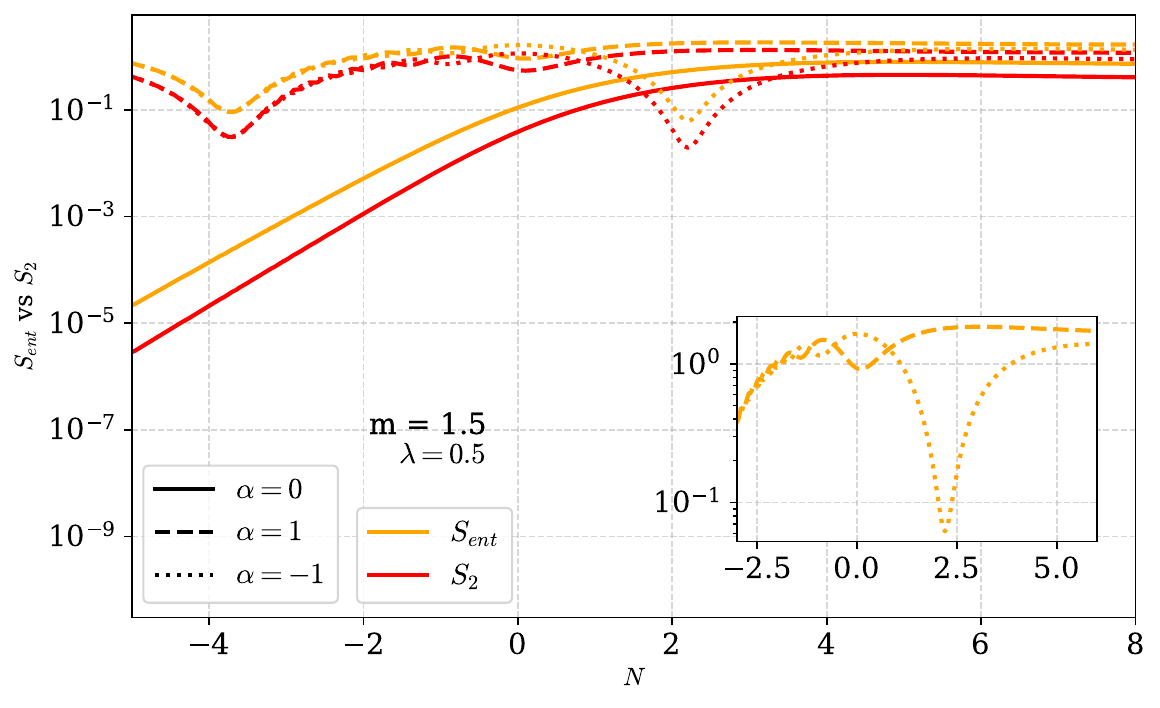}
    \caption{Strong coupling $\lambda=0.5$, $m/H=1.5$. The stronger interaction amplifies pre-crossing modulations and accelerates entropy production after horizon exit; the ordering $S_2<S_{\rm ent}$ holds at all times for all $\alpha$. It is worth noting the small dip after the horizon crossing shared by both the $\alpha =\pm1$ cases. }
    \label{fig:entropy-strong-m15}
\end{figure}

\begin{figure}[t]
    \centering
    \includegraphics[width=0.82\linewidth]{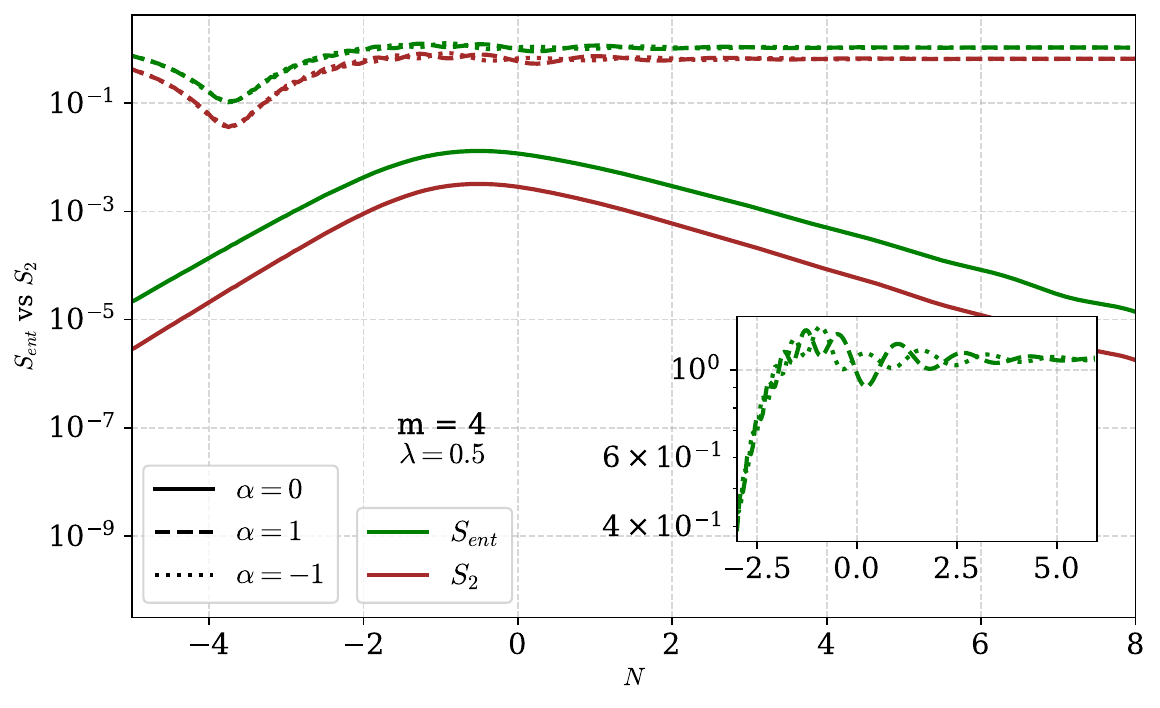}
    \caption{Strong coupling $\lambda=0.5$, $m/H=4$. The BD solutions exhibit a clear decoherence process, while both of the $\alpha = \pm 1 $ show, after a dip inside the horizon, an asymptotic plateau for both the entropy definitions.}
    \label{fig:entropy-strong-m4}
\end{figure}

In Fig.\ref{fig:purity-scan} we display the purity $\gamma(N)$ for a fiducial mode crossing the horizon at $N_{\rm cr}=0$, scanning the mass of the environment $m/H = {1,1.5,4}$ and the initial system state $\alpha\in\{-1,0,1\}$, for weak coupling, $\lambda=0.05$, and strong coupling, $\lambda=0.5$. We complement this with Rényi-2 entropy $S_2=-\ln\gamma$ and von Neumann entropy $S_{\rm ent}$ at fixed $(m,\lambda)$ to cross-check monotonicity, bounding relations, and late-time trends in Figs. \ref{fig:entropy-weak-m1}--\ref{fig:entropy-strong-m4}, with insets zooming pre- and near-horizon regions where oscillatory features or non-monotonicities are more visible. 

\paragraph{Weak coupling $\lambda=0.05$.}
Let's focus on the Fig. \ref{fig:purity-scan} (a). For $m/H=1$, the BD case $\alpha=0$ exhibits a clear decoherence with $\gamma$ monotonically decreasing after horizon exit, without a late-time recovery; Fig. \ref{fig:purity-scan}(a) confirms the absence of a super-horizon rebound and shows a smooth approach to very small $\gamma$ at large $N$. For $\alpha=+1$, a shallow super-horizon uptick appears shortly after $N\simeq 0$ (a few e-folds), followed by a delayed but complete decoherence with $\gamma\to 0$ at very late times; $\alpha=-1$ tracks $\alpha=+1$ with only minor phase-sensitive differences.

For $m/H=1.5$, the BD trajectory $\alpha=0$ shows the purity freezing behaviour. In the Bunch Davies case, the purity is basically frozen to the value $\gamma =1$ until it exits the horizon, when it makes a small drop below the unity, freezing soon after.

For $\alpha=\pm 1$,  there is a slight difference, particularly in the sub-horizon regime. While in the benchmark case the quantum state remains pure until the horizon crossing, the presence of an excitation for the initial system's state induces a primordial drop as it is resolved in Fig. \ref{fig:purity-scan} (a). While both of the cases show a purity freezing behaviour at a late time, the initial condition sets the plateau at which the purity freezes. It is worth noting that after the horizon crossing, the curve for $\alpha = -1$ seems to be above the one with $\alpha = +1$, but after a short transient, the hierarchy is reverted. 

For $m/H=4$ (heavy environment \footnote{It is worth nothing that this also correnspond to a case when the index $\nu_E$ gets an immaginary value.}), the BD curve $\alpha=0$ recovers the non-monotonic behavior highlighted in \cite{Colas:2022kfu}: after an initial decay, $\gamma$ turns upward near and after horizon crossing and heads back toward unity, signaling recoherence in the vacuum case; this can be seen more clearly from the entropy patterns in Fig. \ref{fig:entropy-weak-m4} and Fig. \ref{fig:entropy-strong-m4}. This trend matches the benchmark behaviour and validates the numerical setup. For $\alpha=\pm 1$, initial squeezing introduces phase-sensitive correlations that generate a visible super-horizon rebound in Fig. \ref{fig:purity-scan}, but the late-time trend reverses and $\gamma$ ultimately freezes at a new plateau, confirming that excitations suppress the BD recoherence and lead to a loss of quantumness even in the heavy-environment regime. Notice that for this choice of $m$, given the chosen scale on the vertical axis, the evolution for the cases $\alpha=\pm 1$ are barely distinguishable, but it is worth noting that, regardless of the value of the $\alpha$ excitation, the plateau is the same anyway.

\paragraph{Strong coupling $\lambda=0.5$.}
Increasing the coupling by a factor ten produces oscillations in $\gamma$ while the mode is still sub-horizon in the strong-coupling panel of Fig. \ref{fig:purity-scan}(b), a clear signature of stronger coherent exchange between system and environment prior to crossing; the oscillation amplitude and phase depend on $m/H$, with heavier environments showing larger modulations in Fig. \ref{fig:purity-scan}. 

For $m/H=1$, all $\alpha$ display a faster net loss of purity after crossing compared to the weak-coupling case in Fig. \ref{fig:purity-scan}(a), with $\alpha=\pm 1$ again showing a small super-horizon rise before an eventual fall toward $\gamma\to 0$ at late times. We can note a more pronounced jump for the non-Bunch-Davies curves soon after they cross the horizon, but the overall effect in this case is just to accelerate the process of decoherence, as can be seen from the Fig. \ref{fig:purity-scan}(b). 

For $m/H = 1.5$ we still have the phenomenon of purity freezing, but in this case, the hierarchy between the $\alpha = \pm 1 $ is reversed.
For $m/H=4$, the BD trajectory retains the non-monotonic pattern  For $\alpha=\pm 1$, the stronger coupling enhances the pre-crossing oscillations and the early super-horizon bump in Fig. \ref{fig:purity-scan}, yet the asymptotic trend remains a complete freezing of the purity functions, preventing in all the cases a pure recoherence.

\paragraph{Approximate symmetry $\alpha\leftrightarrow-\alpha$}
A distinctive feature emerges when a massive environment ($m/H = 4$) interacts with an excited system. While a non-zero parameter $\alpha$ determines the plateau values in both weak and strong coupling regimes, the case $m/H = 4$ exhibits a notable symmetry: the sign of $\alpha$ does not affect the asymptotic behaviour at late times. Instead, the plateau value in both panels of figure \ref{fig:purity-scan} depend solely on the coupling strength $\lambda$, with both signs of $\alpha$ yielding identical late-time dynamics despite different transient evolution.

\paragraph{Rényi-2 vs von Neumann entropies.}
At fixed $(m,\lambda)$, $S_2$ lies below $S_{\rm ent}$ at all times and tracks its time dependence in Figs. \ref{fig:entropy-weak-m1}--\ref{fig:entropy-strong-m4}, validating $S_2$ as a robust proxy for information flow and mixedness in the Gaussian setting. For BD and heavy environment ($m/H=4$), we find the characteristic non-monotonicity shown in Fig. \ref{fig:entropy-weak-m4} and Fig. \ref{fig:entropy-strong-m4}: after an initial rise (decoherence) both entropies turn down and approach near-zero values consistent with recoherence, in agreement with the purity analysis and the vacuum benchmark. A comparison with the corresponding purities in Fig. 1 shows how the logarithmic scaling of the entropies enhances the visibility of the recoherence pattern.

For $\alpha=\pm 1$, both entropies exhibit a small super-horizon dip (reflecting the transient purity rebound) followed by a renewed growth that persists to late times, as visible in Figs. \ref{fig:entropy-weak-m1}--\ref{fig:entropy-weak-m15} and Figs. \ref{fig:entropy-strong-m1}--\ref{fig:entropy-strong-m4}; the strong-coupling case enhances early oscillations and slightly increases the entropy production rate after horizon exit, as can be seen by comparing Fig. \ref{fig:entropy-weak-m1} vs Fig. \ref{fig:entropy-strong-m1} and Fig. \ref{fig:entropy-weak-m15} vs Fig. \ref{fig:entropy-strong-m15}. The ordering $S_2 < S_{\rm ent}$ is preserved throughout in all panels of Figs. \ref{fig:entropy-weak-m1}--\ref{fig:entropy-strong-m4}, and late-time growth rates are controlled primarily by $(\lambda,m/H)$. In all the cases, as it can be seen by the insets, around the horizon crossing, transient non-monotonic features show up.


\section{Conclusions}
\label{sec:conclusions}
We investigated the quantum-to-classical transition of primordial perturbations in a two-field inflationary setup where an adiabatic system interacts with an entropic environment through a derivative mixing. Departing from the standard Bunch--Davies assumption for the system, we initialized the adiabatic mode in Gaussian excited states parameterized by a squeezing amplitude $\alpha$, while keeping the environment in BD. We evolved the full $4\times4$ phase-space covariance with the Transport Method, monitoring purity $\gamma$, Rényi-2 entropy $S_2$, and von Neumann entropy $S_{\rm ent}$ across horizon crossing for representative masses $m/H$ and couplings $\lambda$. Our analysis shows that the initial excitation of the system qualitatively alters late-time coherence. For $\alpha=0$ (BD system), we reproduce the known non-monotonic behaviour for heavy environments ($m/H\simeq 4$), where a post-crossing upturn drives $\gamma$ back toward unity and both $S_2$ and $S_{\rm ent}$ toward near-zero values, in line with the vacuum recoherence benchmark. In contrast, for $\alpha\neq 0$ we find a characteristic pattern: a small super-horizon rebound of $\gamma$ (visible in the insets for the strong-coupling regime) followed by a purity freezing at very late times, with $\gamma$ tending to different plateaus and entropies increasing again. This indicates that initial phase-sensitive correlations encoded by squeezing parameters crucially impact the late-time bahaviour of the quantumness of the system.

Stronger coupling $\lambda=0.5$ induces visible sub-horizon oscillations in $\gamma$ and enhances early-time modulations in $S_2$ and $S_{\rm ent}$, reflecting more efficient pre-crossing information exchange. Post-crossing entropy production is accelerated relative to the weak-coupling case, yet the qualitative dichotomy persists: vacuum recoherence appears only for heavy environments and $\alpha=0$, whereas any $\alpha\neq 0$ blocks a complete decoherence. The insets across figures isolate the near-horizon non-monotonic segments and confirm the consistency of our implementation with the established vacuum baseline.

The present results suggest that excited initial conditions can suppress vacuum recoherence and leave residual signatures in phase-sensitive observables. 
This, once again, places the role of the attractor vacuum (the Bunch-Davies one) as a very special case over all the possible invariant de Sitter states.

In the future it would be interesting to extend the analysis to $k$-dependent excitation profiles $\alpha_k$, slow-roll drift of $H$, and non-attractor backgrounds, which will allow assessing robustness and potential observational imprints, including corrections to the power spectrum and to higher-order statistics. 

\vspace{0.5truecm}
{\bf Note.} Authors order reflects their relative contributions to this work.

\acknowledgments
Mattia Cielo thanks Jaime Calderón-Figueroa for helpful conversations. 
All authors are supported by Ministero dell’Universit\`a e della Ricerca (MUR), PRIN2022 program (Grant PANTHEON 2022E2J4RK) Italy, and 
by the Research grant TAsP (Theoretical Astroparticle Physics) funded by Istituto Nazionale di Fisica Nucleare (INFN).
For the purpose of open access, the authors have applied a Creative Commons Attribution (CC-BY) licence to any Author Accepted Manuscript version arising from this work.

\appendix

\bibliographystyle{JHEP}
\bibliography{open}

\providecommand{\href}[2]{#2}\begingroup\raggedright\begin{thebibliography}{10}

\bibitem{Starobinsky:1980te}
A.A.~Starobinsky, \emph{{A New Type of Isotropic Cosmological Models Without
  Singularity}},
  \href{https://doi.org/10.1016/0370-2693(80)90670-X}{\emph{Phys. Lett. B}
  {\bfseries 91} (1980) 99}.

\bibitem{Sato:1981ds}
K.~Sato, \emph{{Cosmological Baryon Number Domain Structure and the First Order
  Phase Transition of a Vacuum}},
  \href{https://doi.org/10.1016/0370-2693(81)90805-4}{\emph{Phys. Lett. B}
  {\bfseries 99} (1981) 66}.

\bibitem{PhysRevD.23.347}
A.H.~Guth, \emph{Inflationary universe: A possible solution to the horizon and
  flatness problems},
  \href{https://doi.org/10.1103/PhysRevD.23.347}{\emph{Phys. Rev. D} {\bfseries
  23} (1981) 347}.

\bibitem{Linde:1981mu}
A.D.~Linde, \emph{{A New Inflationary Universe Scenario: A Possible Solution of
  the Horizon, Flatness, Homogeneity, Isotropy and Primordial Monopole
  Problems}}, \href{https://doi.org/10.1016/0370-2693(82)91219-9}{\emph{Phys.
  Lett. B} {\bfseries 108} (1982) 389}.

\bibitem{Lemoine:2008zz}
M.~Lemoine, J.~Martin and P.~Peter, eds., \emph{{Inflationary cosmology}}
  (2008),
  \href{https://doi.org/10.1007/978-3-540-74353-8}{10.1007/978-3-540-74353-8}.

\bibitem{Brandenberger:1990bx}
R.H.~Brandenberger, R.~Laflamme and M.~Mijic, \emph{{Classical Perturbations
  From Decoherence of Quantum Fluctuations in the Inflationary Universe}},
  \href{https://doi.org/10.1142/S0217732390002651}{\emph{Mod. Phys. Lett. A}
  {\bfseries 5} (1990) 2311}.

\bibitem{Polarski:1995jg}
D.~Polarski and A.A.~Starobinsky, \emph{{Semiclassicality and decoherence of
  cosmological perturbations}},
  \href{https://doi.org/10.1088/0264-9381/13/3/006}{\emph{Class. Quant. Grav.}
  {\bfseries 13} (1996) 377}
  [\href{https://arxiv.org/abs/gr-qc/9504030}{{\ttfamily gr-qc/9504030}}].

\bibitem{Kiefer:1998qe}
C.~Kiefer, D.~Polarski and A.A.~Starobinsky, \emph{{Quantum to classical
  transition for fluctuations in the early universe}},
  \href{https://doi.org/10.1142/S0218271898000292}{\emph{Int. J. Mod. Phys. D}
  {\bfseries 7} (1998) 455}
  [\href{https://arxiv.org/abs/gr-qc/9802003}{{\ttfamily gr-qc/9802003}}].

\bibitem{Burgess:2006jn}
C.P.~Burgess, R.~Holman and D.~Hoover, \emph{{Decoherence of inflationary
  primordial fluctuations}},
  \href{https://doi.org/10.1103/PhysRevD.77.063534}{\emph{Phys. Rev. D}
  {\bfseries 77} (2008) 063534}
  [\href{https://arxiv.org/abs/astro-ph/0601646}{{\ttfamily
  astro-ph/0601646}}].

\bibitem{Kranas:2025jgm}
D.~Kranas, J.~Grain and V.~Vennin, \emph{{Recoherence, adiabaticity, and
  Markovianity in Gaussian maps}},
  \href{https://doi.org/10.1103/r8pq-kms4}{\emph{Phys. Rev. D} {\bfseries 112}
  (2025) 076014} [\href{https://arxiv.org/abs/2504.18648}{{\ttfamily
  2504.18648}}].

\bibitem{Lopez:2025arw}
F.~Lopez and N.~Bartolo, \emph{{Quantum signatures and decoherence during
  inflation from deep subhorizon perturbations}},
  \href{https://arxiv.org/abs/2503.23150}{{\ttfamily 2503.23150}}.

\bibitem{Colas:2022hlq}
T.~Colas, J.~Grain and V.~Vennin, \emph{{Benchmarking the cosmological master
  equations}},
  \href{https://doi.org/10.1140/epjc/s10052-022-11047-9}{\emph{Eur. Phys. J. C}
  {\bfseries 82} (2022) 1085}
  [\href{https://arxiv.org/abs/2209.01929}{{\ttfamily 2209.01929}}].

\bibitem{Colas:2022kfu}
T.~Colas, J.~Grain and V.~Vennin, \emph{{Quantum recoherence in the early
  universe}}, \href{https://doi.org/10.1209/0295-5075/acdd94}{\emph{EPL}
  {\bfseries 142} (2023) 69002}
  [\href{https://arxiv.org/abs/2212.09486}{{\ttfamily 2212.09486}}].

\bibitem{Martin:2000xs}
J.~Martin and R.H.~Brandenberger, \emph{{The TransPlanckian problem of
  inflationary cosmology}},
  \href{https://doi.org/10.1103/PhysRevD.63.123501}{\emph{Phys. Rev. D}
  {\bfseries 63} (2001) 123501}
  [\href{https://arxiv.org/abs/hep-th/0005209}{{\ttfamily hep-th/0005209}}].

\bibitem{Martin:2021znx}
J.~Martin, A.~Micheli and V.~Vennin, \emph{{Discord and decoherence}},
  \href{https://doi.org/10.1088/1475-7516/2022/04/051}{\emph{JCAP} {\bfseries
  04} (2022) 051} [\href{https://arxiv.org/abs/2112.05037}{{\ttfamily
  2112.05037}}].

\bibitem{Calzetta_Hu_2008}
E.A.~Calzetta and B.-L.B.~Hu, \emph{Nonequilibrium Quantum Field Theory},
  Cambridge Monographs on Mathematical Physics, Cambridge University Press
  (2008).

\bibitem{Martin_2018_a}
J.~Martin and V.~Vennin, \emph{Observational constraints on quantum decoherence
  during inflation},
  \href{https://doi.org/10.1088/1475-7516/2018/05/063}{\emph{Journal of
  Cosmology and Astroparticle Physics} {\bfseries 2018} (2018) 063–063}.

\bibitem{Martin_2018_b}
J.~Martin and V.~Vennin, \emph{Non gaussianities from quantum decoherence
  during inflation},
  \href{https://doi.org/10.1088/1475-7516/2018/06/037}{\emph{Journal of
  Cosmology and Astroparticle Physics} {\bfseries 2018} (2018) 037–037}.

\bibitem{Brahma_2022_a}
S.~Brahma, A.~Berera and J.~Calderón-Figueroa, \emph{Universal signature of
  quantum entanglement across cosmological distances},
  \href{https://doi.org/10.1088/1361-6382/aca066}{\emph{Classical and Quantum
  Gravity} {\bfseries 39} (2022) 245002}.

\bibitem{Brahma_2022_b}
S.~Brahma, A.~Berera and J.~Calderón-Figueroa, \emph{Quantum corrections to
  the primordial tensor spectrum: open efts \& markovian decoupling of uv
  modes}, \href{https://doi.org/10.1007/jhep08(2022)225}{\emph{Journal of High
  Energy Physics} {\bfseries 2022} (2022) }.

\bibitem{Assassi:2013gxa}
V.~Assassi, D.~Baumann, D.~Green and L.~McAllister, \emph{{Planck-Suppressed
  Operators}}, \href{https://doi.org/10.1088/1475-7516/2014/01/033}{\emph{JCAP}
  {\bfseries 01} (2014) 033} [\href{https://arxiv.org/abs/1304.5226}{{\ttfamily
  1304.5226}}].

\bibitem{Boyanovsky_2015}
D.~Boyanovsky, \emph{Effective field theory during inflation: Reduced density
  matrix and its quantum master equation},
  \href{https://doi.org/10.1103/physrevd.92.023527}{\emph{Physical Review D}
  {\bfseries 92} (2015) }.

\bibitem{Boyanovsky_2016}
D.~Boyanovsky, \emph{Effective field theory during inflation. ii. stochastic
  dynamics and power spectrum suppression},
  \href{https://doi.org/10.1103/physrevd.93.043501}{\emph{Physical Review D}
  {\bfseries 93} (2016) }.

\bibitem{Hollowood:2017bil}
T.J.~Hollowood and J.I.~McDonald, \emph{{Decoherence, discord and the quantum
  master equation for cosmological perturbations}},
  \href{https://doi.org/10.1103/PhysRevD.95.103521}{\emph{Phys. Rev. D}
  {\bfseries 95} (2017) 103521}
  [\href{https://arxiv.org/abs/1701.02235}{{\ttfamily 1701.02235}}].

\bibitem{Burgess:2022nwu}
C.P.~Burgess, R.~Holman, G.~Kaplanek, J.~Martin and V.~Vennin, \emph{{Minimal
  decoherence from inflation}},
  \href{https://doi.org/10.1088/1475-7516/2023/07/022}{\emph{JCAP} {\bfseries
  07} (2023) 022} [\href{https://arxiv.org/abs/2211.11046}{{\ttfamily
  2211.11046}}].

\bibitem{Caldeira:1981rx}
A.O.~Caldeira and A.J.~Leggett, \emph{{Influence of dissipation on quantum
  tunneling in macroscopic systems}},
  \href{https://doi.org/10.1103/PhysRevLett.46.211}{\emph{Phys. Rev. Lett.}
  {\bfseries 46} (1981) 211}.

\bibitem{Eisert_2002}
J.~Eisert and M.B.~Plenio, \emph{Quantum and classical correlations in quantum
  brownian motion},
  \href{https://doi.org/10.1103/physrevlett.89.137902}{\emph{Physical Review
  Letters} {\bfseries 89} (2002) }.

\bibitem{Brahma:2024ycc}
S.~Brahma, J.~Calder{\'o}n-Figueroa, X.~Luo and D.~Seery, \emph{{The special
  case of slow-roll attractors in de~Sitter: non-Markovian noise and evolution
  of entanglement entropy}},
  \href{https://doi.org/10.1088/1475-7516/2025/04/050}{\emph{JCAP} {\bfseries
  04} (2025) 050} [\href{https://arxiv.org/abs/2411.08632}{{\ttfamily
  2411.08632}}].

\bibitem{Brandenberger:2012aj}
R.H.~Brandenberger and J.~Martin, \emph{{Trans-Planckian Issues for
  Inflationary Cosmology}},
  \href{https://doi.org/10.1088/0264-9381/30/11/113001}{\emph{Class. Quant.
  Grav.} {\bfseries 30} (2013) 113001}
  [\href{https://arxiv.org/abs/1211.6753}{{\ttfamily 1211.6753}}].

\bibitem{Ashoorioon:2014nta}
A.~Ashoorioon, K.~Dimopoulos, M.M.~Sheikh-Jabbari and G.~Shiu,
  \emph{{Non-Bunch\textendash{}Davis initial state reconciles chaotic models
  with BICEP and Planck}},
  \href{https://doi.org/10.1016/j.physletb.2014.08.038}{\emph{Phys. Lett. B}
  {\bfseries 737} (2014) 98} [\href{https://arxiv.org/abs/1403.6099}{{\ttfamily
  1403.6099}}].

\bibitem{Kundu:2011sg}
S.~Kundu, \emph{{Inflation with General Initial Conditions for Scalar
  Perturbations}},
  \href{https://doi.org/10.1088/1475-7516/2012/02/005}{\emph{JCAP} {\bfseries
  02} (2012) 005} [\href{https://arxiv.org/abs/1110.4688}{{\ttfamily
  1110.4688}}].

\bibitem{Akama:2020jko}
S.~Akama, S.~Hirano and T.~Kobayashi, \emph{{Primordial tensor
  non-Gaussianities from general single-field inflation with non-Bunch-Davies
  initial states}},
  \href{https://doi.org/10.1103/PhysRevD.102.023513}{\emph{Phys. Rev. D}
  {\bfseries 102} (2020) 023513}
  [\href{https://arxiv.org/abs/2003.10686}{{\ttfamily 2003.10686}}].

\bibitem{BouzariNezhad:2018zsi}
H.~Bouzari~Nezhad and F.~Shojai, \emph{{The Effect of $\alpha$-Vacua on the
  Scalar and Tensor Spectral Indices: Slow-Roll Approximation}},
  \href{https://doi.org/10.1103/PhysRevD.98.063512}{\emph{Phys. Rev. D}
  {\bfseries 98} (2018) 063512}
  [\href{https://arxiv.org/abs/1802.05537}{{\ttfamily 1802.05537}}].

\bibitem{Planck:2018vyg}
{\scshape Planck} collaboration, \emph{{Planck 2018 results. VI. Cosmological
  parameters}},
  \href{https://doi.org/10.1051/0004-6361/201833910}{\emph{Astron. Astrophys.}
  {\bfseries 641} (2020) A6}
  [\href{https://arxiv.org/abs/1807.06209}{{\ttfamily 1807.06209}}].

\bibitem{Starobinsky:1992ts}
A.A.~Starobinsky, \emph{{Spectrum of adiabatic perturbations in the universe
  when there are singularities in the inflation potential}}, {\emph{JETP Lett.}
  {\bfseries 55} (1992) 489}.

\bibitem{Brahma:2024yor}
S.~Brahma, J.~Calder{\'o}n-Figueroa and X.~Luo, \emph{{Time-convolutionless
  cosmological master equations: late-time resummations and decoherence for
  non-local kernels}},
  \href{https://doi.org/10.1088/1475-7516/2025/08/019}{\emph{JCAP} {\bfseries
  08} (2025) 019} [\href{https://arxiv.org/abs/2407.12091}{{\ttfamily
  2407.12091}}].

\bibitem{MUKHANOV1992203}
V.~Mukhanov, H.~Feldman and R.~Brandenberger, \emph{Theory of cosmological
  perturbations},
  \href{https://doi.org/https://doi.org/10.1016/0370-1573(92)90044-Z}{\emph{Physics
  Reports} {\bfseries 215} (1992) 203}.

\bibitem{Mukhanov:2007zz}
V.~Mukhanov and S.~Winitzki, \emph{{Introduction to quantum effects in
  gravity}}, Cambridge University Press (6, 2007).

\bibitem{Cielo:2022vmo}
M.~Cielo, G.~Mangano and O.~Pisanti, \emph{{Impact of trans-Planckian quantum
  noise on the primordial gravitational wave spectrum}},
  \href{https://doi.org/10.1103/PhysRevD.108.043501}{\emph{Phys. Rev. D}
  {\bfseries 108} (2023) 043501}
  [\href{https://arxiv.org/abs/2211.04316}{{\ttfamily 2211.04316}}].

\bibitem{Cielo:2023enz}
M.~Cielo, M.~Fasiello, G.~Mangano and O.~Pisanti, \emph{{Gravitational wave
  non-Gaussianity from trans-Planckian quantum noise}},
  \href{https://doi.org/10.1088/1475-7516/2024/04/079}{\emph{JCAP} {\bfseries
  04} (2024) 079} [\href{https://arxiv.org/abs/2309.12285}{{\ttfamily
  2309.12285}}].

\bibitem{Cielo:2024poz}
M.~Cielo, G.~Mangano, O.~Pisanti and D.~Wands, \emph{{Steepest growth in the
  primordial power spectrum from excited states at a sudden transition}},
  \href{https://doi.org/10.1088/1475-7516/2025/04/007}{\emph{JCAP} {\bfseries
  04} (2025) 007} [\href{https://arxiv.org/abs/2410.22154}{{\ttfamily
  2410.22154}}].

\bibitem{Borde:2001nh}
A.~Borde, A.H.~Guth and A.~Vilenkin, \emph{{Inflationary space-times are
  incompletein past directions}},
  \href{https://doi.org/10.1103/PhysRevLett.90.151301}{\emph{Phys. Rev. Lett.}
  {\bfseries 90} (2003) 151301}
  [\href{https://arxiv.org/abs/gr-qc/0110012}{{\ttfamily gr-qc/0110012}}].

\bibitem{Kempf:2000ac}
A.~Kempf, \emph{{Mode generating mechanism in inflation with cutoff}},
  \href{https://doi.org/10.1103/PhysRevD.63.083514}{\emph{Phys. Rev. D}
  {\bfseries 63} (2001) 083514}
  [\href{https://arxiv.org/abs/astro-ph/0009209}{{\ttfamily
  astro-ph/0009209}}].

\bibitem{Holman:2007na}
R.~Holman and A.J.~Tolley, \emph{{Enhanced Non-Gaussianity from Excited Initial
  States}}, \href{https://doi.org/10.1088/1475-7516/2008/05/001}{\emph{JCAP}
  {\bfseries 05} (2008) 001} [\href{https://arxiv.org/abs/0710.1302}{{\ttfamily
  0710.1302}}].

\bibitem{Brahma:2019unn}
S.~Brahma, \emph{{Trans-Planckian censorship, inflation and excited initial
  states for perturbations}},
  \href{https://doi.org/10.1103/PhysRevD.101.023526}{\emph{Phys. Rev. D}
  {\bfseries 101} (2020) 023526}
  [\href{https://arxiv.org/abs/1910.04741}{{\ttfamily 1910.04741}}].

\bibitem{Ashoorioon:2010xg}
A.~Ashoorioon and G.~Shiu, \emph{{A Note on Calm Excited States of Inflation}},
  \href{https://doi.org/10.1088/1475-7516/2011/03/025}{\emph{JCAP} {\bfseries
  03} (2011) 025} [\href{https://arxiv.org/abs/1012.3392}{{\ttfamily
  1012.3392}}].

\bibitem{Brahma:2025dio}
S.~Brahma and J.~Calder{\'o}n-Figueroa, \emph{{Is the CMB revealing signs of
  pre-inflationary physics?}},
  \href{https://arxiv.org/abs/2504.02746}{{\ttfamily 2504.02746}}.

\bibitem{Gong:2023kpe}
J.-O.~Gong, M.~Mylova and M.~Sasaki, \emph{{New shape of parity-violating
  graviton non-Gaussianity}},
  \href{https://doi.org/10.1007/JHEP10(2023)140}{\emph{JHEP} {\bfseries 10}
  (2023) 140} [\href{https://arxiv.org/abs/2303.05178}{{\ttfamily
  2303.05178}}].

\bibitem{Aravind:2013lra}
A.~Aravind, D.~Lorshbough and S.~Paban, \emph{{Non-Gaussianity from Excited
  Initial Inflationary States}},
  \href{https://doi.org/10.1007/JHEP07(2013)076}{\emph{JHEP} {\bfseries 07}
  (2013) 076} [\href{https://arxiv.org/abs/1303.1440}{{\ttfamily 1303.1440}}].

\bibitem{Kanno:2022mkx}
S.~Kanno and M.~Sasaki, \emph{{Graviton non-gaussianity in
  \ensuremath{\alpha}-vacuum}},
  \href{https://doi.org/10.1007/JHEP08(2022)210}{\emph{JHEP} {\bfseries 08}
  (2022) 210} [\href{https://arxiv.org/abs/2206.03667}{{\ttfamily
  2206.03667}}].

\bibitem{Allen:1985ux}
B.~Allen, \emph{{Vacuum States in de Sitter Space}},
  \href{https://doi.org/10.1103/PhysRevD.32.3136}{\emph{Phys. Rev. D}
  {\bfseries 32} (1985) 3136}.

\bibitem{Danielsson:2002kx}
U.H.~Danielsson, \emph{{A Note on inflation and transPlanckian physics}},
  \href{https://doi.org/10.1103/PhysRevD.66.023511}{\emph{Phys. Rev. D}
  {\bfseries 66} (2002) 023511}
  [\href{https://arxiv.org/abs/hep-th/0203198}{{\ttfamily hep-th/0203198}}].

\bibitem{Broy:2016zik}
B.J.~Broy, \emph{{Corrections to $n_s$ and $n_t$ from high scale physics}},
  \href{https://doi.org/10.1103/PhysRevD.94.103508}{\emph{Phys. Rev. D}
  {\bfseries 94} (2016) 103508}
  [\href{https://arxiv.org/abs/1609.03570}{{\ttfamily 1609.03570}}].

\bibitem{Seery:2012vj}
D.~Seery, D.J.~Mulryne, J.~Frazer and R.H.~Ribeiro, \emph{{Inflationary
  perturbation theory is geometrical optics in phase space}},
  \href{https://doi.org/10.1088/1475-7516/2012/09/010}{\emph{JCAP} {\bfseries
  09} (2012) 010} [\href{https://arxiv.org/abs/1203.2635}{{\ttfamily
  1203.2635}}].

\bibitem{Boyanovsky_2018}
D.~Boyanovsky, \emph{Imprint of entanglement entropy in the power spectrum of
  inflationary fluctuations},
  \href{https://doi.org/10.1103/physrevd.98.023515}{\emph{Physical Review D}
  {\bfseries 98} (2018) }.

\bibitem{PhysRevLett.109.190502}
G.~Adesso, D.~Girolami and A.~Serafini, \emph{Measuring gaussian quantum
  information and correlations using the r\'enyi entropy of order 2},
  \href{https://doi.org/10.1103/PhysRevLett.109.190502}{\emph{Phys. Rev. Lett.}
  {\bfseries 109} (2012) 190502}.

\bibitem{Adesso_2012}
G.~Adesso, D.~Girolami and A.~Serafini, \emph{Measuring gaussian quantum
  information and correlations using the rényi entropy of order 2},
  \href{https://doi.org/10.1103/physrevlett.109.190502}{\emph{Physical Review
  Letters} {\bfseries 109} (2012) }.

\bibitem{Baez:2011upp}
J.C.~Baez, \emph{{R{\'e}nyi Entropy and Free Energy}},
  \href{https://doi.org/10.3390/e24050706}{\emph{Entropy} {\bfseries 24} (2022)
  706} [\href{https://arxiv.org/abs/1102.2098}{{\ttfamily 1102.2098}}].

\bibitem{Kirchanov:2008}
V.~Kirchanov, \emph{Using the renyi entropy to describe quantum dissipative
  systems in statistical mechanics},
  \href{https://doi.org/10.1007/s11232-008-0111-y}{\emph{Theoretical and
  Mathematical Physics} {\bfseries 156} (2008) 1347}.

\bibitem{Lavagno_2002}
A.~Lavagno, \emph{Relativistic nonextensive thermodynamics},
  \href{https://doi.org/10.1016/s0375-9601(02)00964-7}{\emph{Physics Letters A}
  {\bfseries 301} (2002) 13–18}.

\bibitem{kusuki2025universalityrenyientropyconformal}
Y.~Kusuki, H.~Ooguri and S.~Pal, \emph{Universality of r\'enyi entropy in
  conformal field theory},  2025.

\bibitem{Kudler_Flam_2023}
J.~Kudler-Flam, \emph{Rényi mutual information in quantum field theory},
  \href{https://doi.org/10.1103/physrevlett.130.021603}{\emph{Physical Review
  Letters} {\bfseries 130} (2023) }.

\bibitem{Tsallis_2013}
C.~Tsallis and L.J.L.~Cirto, \emph{Black hole thermodynamical entropy},
  \href{https://doi.org/10.1140/epjc/s10052-013-2487-6}{\emph{The European
  Physical Journal C} {\bfseries 73} (2013) }.

\bibitem{abreu2025modifiedtsallisrenyientropymondlike}
E.M.C.~Abreu and J.A.~Neto, \emph{From modified tsallis-renyi entropy to a
  mond-like force law, bekenstein bound, and landauer principle for black
  holes},  2025.

\bibitem{Czinner_2025}
V.G.~Czinner and H.~Iguchi, \emph{Hawking–rényi thermodynamics of rotating
  black holes from locally kiselev-type behavior},
  \href{https://doi.org/10.1140/epjc/s10052-025-14756-z}{\emph{The European
  Physical Journal C} {\bfseries 85} (2025) }.

\bibitem{Brydges_2019}
T.~Brydges, A.~Elben, P.~Jurcevic, B.~Vermersch, C.~Maier, B.P.~Lanyon et~al.,
  \emph{Probing rényi entanglement entropy via randomized measurements},
  \href{https://doi.org/10.1126/science.aau4963}{\emph{Science} {\bfseries 364}
  (2019) 260–263}.

\bibitem{ozawa2024perspectivephysicalinterpretationsrenyi}
M.~Ozawa and N.~Javerzat, \emph{Perspective on physical interpretations of
  r\'enyi entropy in statistical mechanics},  2024.
\newblock https://doi.org/10.1209/0295-5075/ad5d89.

\bibitem{Elben_2019}
A.~Elben, B.~Vermersch, C.F.~Roos and P.~Zoller, \emph{Statistical correlations
  between locally randomized measurements: A toolbox for probing entanglement
  in many-body quantum states},
  \href{https://doi.org/10.1103/physreva.99.052323}{\emph{Physical Review A}
  {\bfseries 99} (2019) }.

\end{thebibliography}\endgroup

\end{document}